# Let's Get Digital: The many ways the fourth industrial revolution is reshaping the way we think about quality

Nicole Radziwill

The technology landscape is richer and more promising than ever before. In many ways, cloud computing, big data, virtual reality (VR), augmented reality (AR), blockchain, additive manufacturing, artificial intelligence (AI), machine learning (ML), Internet Protocol Version 6 (IPv6), cyber-physical systems and the Internet of Things (IoT) all represent new frontiers. These technologies can help improve product and service quality, and organizational performance.

In many regions, the internet is now as ubiquitous as electricity. Components are relatively cheap. And, a robust ecosystem of open-source software libraries means that engineers can solve problems 100 times faster than just two decades ago.

This digital transformation is leading us toward connected intelligent automation: smart, hyperconnected agents deployed in environments where humans and machines cooperate—and leverage data—to achieve shared goals. This isn't the world's first industrial revolution. In fact, it is its fourth—and the disruptive changes it will bring suggest we'll need a fresh perspective on quality to adapt to it.

**Quality 4.0: A fresh perspective**

Quality 4.0 is the name given to the pursuit of performance excellence during these times of potentially disruptive digital transformation. It comes from "Industry 4.0"—a term coined at Hannover Fair in 2011 to describe the fourth industrial revolution.[1]

That event emphasized the increasing intelligence and interconnectedness of smart manufacturing systems. It reflected on the newest technological innovations, placing them in historical context and tracing the development of key technologies from the 1700s to the present.

During the first industrial revolution (late 1700s and early 1800s), innovations in steam and water power made it possible for production facilities to scale up and expand potential production locations. Before then, manufacturing facilities had to be constructed along rivers so that waterwheels could be used to generate power.

By the late 1800s, the discovery of electricity and development of infrastructure enabled engineers to develop machinery for mass production. Iron ore production increased, enabling machines themselves to be mass produced. In the United States, the expansion of railways made it easier to obtain supplies and deliver finished goods.

The widespread availability of reliable power sparked a renaissance in computing. Toward the end of World War II, digital computing started to emerge from its analog roots. The third industrial revolution came at the end of the 1960s with the invention of the programmable logic controller. This made it possible to automate processes, such as filling and reloading tanks, turning engines on and off, and controlling sequences of events based on the state of the process and changing environmental conditions.

The World Economic Forum (WEF) has been keenly interested in these changes. In 2015, it launched the Digital Transformation Initiative to coordinate research that would help anticipate the effects of these changes on business and society. WEF recognized that we've been actively experiencing digital transformation since the emergence of digital computing in the 1950s: first with mainframes, then client-server computing and PCs, followed by the advent of the internet and early e-commerce sites.

Mobile devices and cloud computing led to a convergence of services, as multiple customer touch points (phone, fax, web and tablets) gradually blended into the single view of the customer that most organizations now have. Just 20 years ago, organizations were barely able to link your phone calls to customer service, emails, and web form queries. Now, it's taken for granted.

The first industrial revolution was characterized by steam-powered machines, and the second by electricity and assembly lines. Innovations in computing and industrial automation defined the third. The fourth industrial revolution brings us machine intelligence, pervasive computing, affordable storage and robust connectivity. How can we leverage them to improve quality and performance?

**Emerging trends in quality**

Because the cost of enabling technologies (such as sophisticated sensors, intelligent algorithms and the computing power to leverage them) has decreased so much over the past decade, organizations can now begin making them part of their digital strategy.

The process of digital transformation is revealing changes in how we perceive customers and organizational boundaries. Organizations no longer will be defined solely by their employees and business partners, but also by the customers who participate—without even explicitly being aware of their integral involvement—in ongoing dialogues that shape the evolution of product lines and new services.

New business models won't necessarily rely on ownership, consumption, centralized production of products or centralized provision of services. The value-based approach will accentuate the importance of trust, transparency and security. New technologies (such as blockchain, as defined in the sidebar "Quality 4.0 Tools") will help us implement and deploy systems to support those changes.

Even though the term "Quality 4.0" hadn't been used yet, the quality implications of the fourth industrial revolution were first described as early as 2015 in the ASQ Future of Quality Report.[2] The study aimed to characterize the evolution of the quality landscape over the next five to 10 years, to prepare the quality community for the challenges of the future.

The authors described how the health and viability of the entire industrial ecosystem would become everyone's concern. They anticipated the emergence of several new perspectives, including:

- A shift of emphasis in the quality profession from efficiency, effectiveness and satisfaction to continuous learning and adaptation.

- Changing boundaries in and between organizations, and how information is shared between different areas, due to information availability and transparency ("shifting seams and transitions").

- Supply chain omniscience and visibility into production processes (being able to monitor and respond to any element in real time).

- An increased emphasis on customer experience, participative markets (in which customers consume and produce energy) and prosumerism (in which customers participate in the design and development of the products they want).

Each of these shifts must be addressed in Quality 4.0. As a result, the way we approach and tackle problems will evolve. For example, we must learn how to manage data over the lifetime of the data rather than that of the organization that collects it. Our concept of voice of the customer will expand, and we'll find ways to listen to the voice of things as well because we'll be able to learn about our customers from the connected objects around them.[3]

**Why now?**

Although the growth and expansion of the internet accelerated innovation in the late 1990s and 2000s, only now are we poised for the fourth industrial revolution. What's changing?

- **Production and availability of information:** More information is available because people and devices are producing it at greater rates than ever before. Falling costs of enabling technologies, such as sensors and actuators, are catalyzing innovation in these areas.

- **Connectivity:** First and foremost, the introduction of IPv6—which defines how data are sent from one computer to another—has ensured that there will be enough addresses to locate the billions of devices that are expected to connect to the internet. The information produced by these devices will be instantly accessible over the internet.

    In addition, improved network infrastructure is expanding the extent of connectivity, making it more widely available and robust. (And unlike the 1980s and 1990s, there are far fewer communications protocols that are commonly encountered, so it's a lot easier to get one device to talk to another device on your network.)

- **Intelligent processing:** Affordable computing capabilities (and processing power) are available to analyze and interpret that information so it can be incorporated into decision-making. High-performance software libraries for advanced processing and visualization of data are easy to find and, in many cases, easy to use. (In the past, software developers had to write their own code for even common tasks. Now, they can use open-source solutions that are battle tested by many.)

- **New modes of interaction:** The ways in which we acquire and interact with information also are changing. In particular, new interfaces, such as AR and VR expand possibilities for training and navigating a hybrid physical-digital environment with greater ease.

- **New modes of production:** 3-D printing, nanotechnology and gene editing are poised to change the nature and means of production in several industries. Technologies for augmenting or enhancing human performance (exoskeletons, brain-computer interfaces and even autonomous vehicles, for example) also will open new mechanisms for innovation in production and distribution. New technologies, such as blockchain, have the potential to change the nature of production as well by challenging ingrained centralized perceptions of trust, control, consensus and value creation.

**Discovery: The new role of quality**

What we recognize as today's quality profession began during the middle of the second industrial revolution, with the methods of scientific management introduced by Henri Fayol in France and Frederick Winslow Taylor in the United States. Factories needed methods to ensure assembly lines ran smoothly—that they produced artifacts to specification, that workers knew how to engage in the production process and that costs were controlled.

As industrial production matured, those methods grew to encompass the design of processes built to produce to specification. In the 1980s and 1990s, the adoption of personal computing once again changed the landscape. Organizations regrouped their quality efforts around the value of culture and active engagement in quality—and total quality management (TQM), lean and Six Sigma gained in popularity.

As connected, intelligent and automated systems are more widely adopted, we can once again expect a renaissance in quality tools and methods. The progression can be summarized through four themes:

1. **Quality as inspection:** In the early days, quality assurance relied on inspecting bad quality out of the total items produced. Walter A. Shewhart's methods for statistical process control helped operators determine whether variation was due to random or special causes.

2. **Quality as design:** Inspired by W. Edwards Deming's recommendation to cease dependence on inspection, more holistic methods emerged for designing quality into processes to prevent quality problems before they occurred.

3. **Quality as empowerment:** TQM and Six Sigma advocate a holistic approach to quality, making it everyone's responsibility and empowering individuals to contribute to continuous improvement.

4. **Quality as discovery:** In an adaptive, intelligent environment, quality depends on how quickly we can discover and aggregate new data sources, how effectively we can discover root causes and how well we can discover new insights about ourselves, our products and our organizations.

**Quality 4.0 value propositions**

How can Quality 4.0 help your organization? Specifically, how can you improve the performance of your people, projects and products by implementing enabling technologies such as AI, ML, robotic process automation and blockchain?

New technology always should be introduced with a clear articulation of its desired benefits. A value proposition is a statement that explains what benefits a product or activity will deliver and, sometimes, how it will happen. Value propositions for Quality 4.0 initiatives fall into six categories, with No. 1 being the most significant:

1. Augment (or improve upon) human intelligence.

2. Increase the speed and quality of decision-making.

3. Improve transparency, traceability and auditability.

4. Anticipate changes, reveal biases and adapt to new circumstances and knowledge.

5. Evolve relationships, organizational boundaries and concept of trust to reveal opportunities for continuous improvement and new business models.

6. Learn how to learn by cultivating self-awareness and other-awareness as skills.

Quality 4.0 initiatives might help you add intelligence to monitoring and managing operations, or enable remote monitoring to improve the productivity or morale of your operators. Think about how to add to human capabilities rather than how to replace people in your processes. According to Finnish sociologist Esko Kilpi, "The real future of work is not in the industrial model of pursuit automation but in the post-industrial model of promoting augmentation."[4]

Predictive maintenance can help you anticipate equipment failures and proactively reduce downtime. Quality 4.0 initiatives can help you assess supply chain risk on an ongoing basis, or help you decide whether to take corrective action. Quality 4.0 initiatives also can help you improve cybersecurity: documenting and benchmarking processes can help your organization detect anomalies and understand expected performance to more effectively flag potential attacks.

**The ecosystem of enabling technologies: How to add value**

Automation isn't an all-or-nothing prospect. A user can create a process that a computer or intelligent agent executes, the computer can make decisions for an operator to approve or adjust, or the computer can make and execute all decisions.

Similarly, machine intelligence is a spectrum: An algorithm can provide advice, take action with approvals or adjustments, or take action on its own. We must decide what value we want to generate when we introduce various degrees of intelligence and automation into our processes.

To do this, we must understand the techniques and technologies of Quality 4.0, as defined in the sidebar "Quality 4.0 Tools," and how they relate to each other. The relationships among these areas are illustrated in Figure 1.

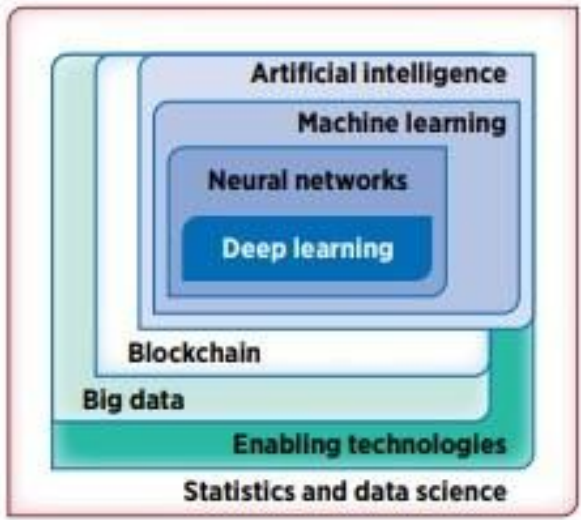

**SIDEBAR: Quality 4.0 Technologies**

- **Artificial intelligence:** computer vision, language processing, chatbots, personal assistants, navigation, robotics, making complex decisions.

- **Big data:** infrastructure (such as MapReduce, Hadoop, Hive and NoSQL databases), easier access to data sources, tools for managing and analyzing large datasets without having to use supercomputers.

- **Blockchain:** increasing transparency and auditability of transactions (for assets and information), monitoring conditions so transactions don't occur unless quality objectives are met.

- **Deep learning:** image classification, complex pattern recognition, time series forecasting, text generation, creating sound and art, creating fictitious video from real video, adjusting images based on heuristics (make a frowning person in a photo appear to smile, for example).

- **Enabling technologies:** affordable sensors and actuators, cloud computing, open-source software, AR, mixed reality, VR, data streaming (such as Kafka and Storm), 5G networks, IPv6, IoT.

- **Machine learning:** text analysis, recommendation systems, email spam filters, fraud detection, classifying objects into groups, forecasting.

- **Data science:** the practice of bringing together heterogeneous data sets for making predictions, performing classifications, finding patterns in large datasets, reducing large sets of observations to most significant predictors, applying sound traditional techniques (such as visualization, inference and simulation) to generate viable models and solutions.

AI encompasses the most tools, which, in many cases, can become so ordinary that they're no longer considered AI, such as optical character recognition. ML algorithms make up some (but not all) of the domain of AI. Neural networks are one kind of ML algorithm, and deep learning is a special kind of complex neural network that incorporates layers with special functions.

AI and ML are becoming popular now because not only is the software more accessible and easier to apply, it's easier to access the big data that makes AI and ML so powerful.

Blockchain—a newly emerging technology—has the potential to improve data quality and the quality of transactions. Statistics and data science provide the firm foundations that should be applied to all problem solving.

**Quality professionals: Leading the transformation**

The introduction of AI and ML means that data-driven decision-making can become more self-aware. With better information, we'll be better able to adapt to changing environments and changing customer or stakeholder needs.

Quality professionals are perfectly positioned to lead digital transformation efforts because we have deep skills in:

- Systems thinking.
- Data-driven decision-making.
- Leadership for organizational learning.
- Establishing processes for continuous improvement.
- Understanding how decisions affect people: lives, relationships, communities, well-being, health and society in general.

This last skill is particularly important. Many ML algorithms must be trained, and training is subject to personal and cognitive biases. Quality professionals can anticipate positive and negative effects, helping organizations protect against negative consequences while capturing opportunities that will benefit everyone.

Quality professionals are distinctively good at structured problem solving, data-driven decision-making and leveraging cultural change to facilitate improvement. In Quality 4.0, these fundamentals will not change, even as the amount and variety of data increase. As a community, we are uniquely positioned to help our organizations thrive in this new era.